\documentclass[
    aps,prl,floatfix,twocolumn,amsfonts,amssymb,amsmath
    ]{revtex4-1} 

\usepackage{dsfont}
\usepackage{float} 
\usepackage[colorlinks=true, linkcolor=black, citecolor=black, filecolor=black,
urlcolor=black, breaklinks=true]{hyperref}
\usepackage{mathtools}
\usepackage[capitalize]{cleveref}
\usepackage{physics}
\usepackage{multibib}
\usepackage{tikz}
\usepackage{xcolor}
\usepackage{csquotes}
\usepackage{natbib}
\usepackage{graphicx} 

\usepackage{siunitx}

\newcommand{\m}{\mathrm}

\begin{document}
\title{Boosting information transfer in a quantum correlated medium}
\author{Finn Schmolke}
\author{Etienne Springer}
\author{Eric Lutz}
\affiliation{Institute for Theoretical Physics I, University of Stuttgart,
D-70550 Stuttgart, Germany}

\begin{abstract}
Sharing and receiving information plays a pivotal role in science and technology. Quantum communication relies on the principles of quantum mechanics to transmit information in a nonclassical manner. Existing quantum communication protocols are commonly based on shared entangled states between sender and receiver, while the transmitting medium is classical.  We here demonstrate that information transfer may be enhanced in a quantum correlated medium without entanglement distribution. We concretely show that  nonclassical correlations, with nonzero discord, between the first two spins of a spin chain  that acts as a quantum wire can increase the information flow  and reduce the propagation time. We relate this effect to the breaking of the spatial symmetry of the out-of-time-order correlator that characterizes the spread of  information through the  medium.
\end{abstract}

\maketitle

Communication  is based on the transfer of information between one point and another
\cite{bri65,ger00,yu01,fra18}. A notable observation is that both communication and information are  inherently physical, since information is transmitted by physical means, through a physical medium, for instance, in the form of an electric current or a light wave \cite{bri65,ger00,yu01,fra18}. They are therefore subjected to the laws of physics, in particular, to those of quantum theory. Quantum mechanics has been shown to restrict  the flow of entropy/information, that is, the number of bits sent per unit time \cite{leb66,bow67,pen83,bek88,bek90,yue92,cav94,ble00}. For a single  channel, the entropy flux $\dot I$ is  upper bounded by the energy current $\dot E$ according to $\dot I^2 \leq {\pi  \dot E/3 \hbar}$ \cite{leb66,bow67,pen83,bek88,bek90,yue92,cav94,ble00}. There is hence a minimum energy cost per bit associated with communication. This bound is believed to be independent of the physical properties of the transmitting medium, and thus universal \cite{ble00}. It  is intimately related to the existence of quantum speed limits that impose strong constraints on the evolution time of any quantum system \cite{def17}. 

We here show that   nonclassical correlations may boost entropy currents in a medium.  Quantum physics, thus, not only fundamentally constrains  information flow, it can    also enhance it. We focus in the following on information transfer along a  spin chain that acts as a quantum wire \cite{bos23,chr04,fit06,kay07,cap07,fra08,ban11,yao11,god12,yao12,ajo13,sah15,mar16,bos07,nik14,cha16}. Whereas photons are ideal carriers of information over long distances, it is not straightforward to convert information between two physically different  photonic qubits. Spin chains are a prominent, solid-state alternative for short- and mid-range communication, for example, to connect quantum processors or quantum registers \cite{bos23,chr04,fit06,kay07,cap07,fra08,ban11,yao11,god12,yao12,ajo13,sah15,mar16,bos07,nik14,cha16}. In these systems,  information is transmitted dynamically from one end of a chain to the other, without the requirement for any external control. We consider for concreteness the perfect state transfer protocol introduced in Ref.~\cite{chr04}, and implemented experimentally in Ref.~\cite{cha16}, where the magnetic interactions between neighboring spins are engineered such that information is transferred with unit fidelity. We demonstrate that the transmission of  a qubit end-state along the  spin chain is  
significantly improved in the presence of nonclassical correlations, with nonzero discord \cite{oll01,hen01,mod12,ber17,hu18}, between only the first two spins of the chain, that is, sender and receiver are initially  uncorrelated. This enhancement is associated with a reduced arrival time and a larger value of the maximum entropy flux. We additionally provide a sufficient condition for this increase of  communication rate  to occur, and illustrate it with a general class of two-qubit $X$ states that can be treated analytically \cite{yu07,ali10,che11,que12}. It is worth emphasizing that this quantum improvement  strongly differs from usual quantum communication protocols, such as quantum key distribution, quantum teleportation and super-dense coding, that usually rely on a shared entangled state between sender and receiver \cite{gis07}. 

\begin{figure*}[t]
  \centering
    \begin{tikzpicture}
      \node (a) [label={[label distance=-0.2 cm]143: \textbf{(a)}}] at (0,0) {\includegraphics[width=0.52\textwidth]{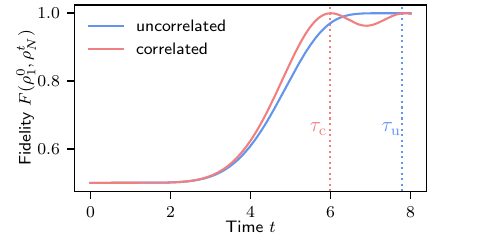}};	
      \node (c) [label={[label distance=-0.2 cm]143: \textbf{(c)}}] at (0,-4.7) {\includegraphics[width=0.52\textwidth]{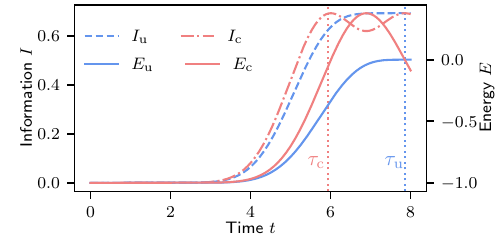}};	
      \node (b) [label={[label distance=-0.2 cm]143: \textbf{(b)}}] at (9.3,0) {\includegraphics[width=0.52\textwidth]{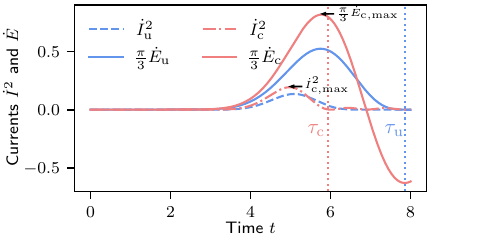}}; 
      \node (d) [label={[label distance=-0.2 cm]143: \textbf{(d)}}] at (9.3,-4.7) {\includegraphics[width=0.52\textwidth]{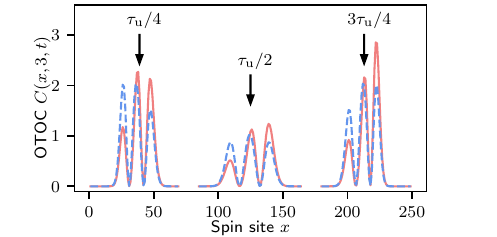}}; 
      \draw (d) node[yshift=2.06cm,xshift=-3.70cm] {\scriptsize $\times 10^{-2}$};
    \end{tikzpicture}
    \caption{Quantum enhanced information transfer along an $XY$ spin chain. a) Perfect state transfer with unit fidelity $F(\rho^{0}_1,\rho^t_N)$ between the edge spins, 1 and $N$, happens faster  with local quantum correlations between the first two spins (time $\tau_\m{c}$, red) than without  (time $\tau_\m{u}$, blue). b) The  information-energy flow inequality, $\dot I^2 \leq {\pi  \dot E/3}$, is obeyed in both cases, and the maximum currents,  ${\dot{I}}_\text{max}$ and $ \dot{E}_\text{max}$, are increased by the nonclassical correlations (arrows). c) In the absence of correlations, information and energy of the receiving qubit increase monotonically to the values of the sender qubit, $I= \ln 2$ and $E = 0$, at the same time $\tau_\m{u}$. In the presence of  correlations, information and energy reach their maxima more rapidly. However, energy is delayed and overshoots the sender value before attaining it at time  $\tau_\m{u}$, after reflection at the end of the chain. d) Out-of-time-order correlator $C(x,3,t)$ that characterizes the propagation of information between  spins 3 and $x$ at three different times, $\tau_\m{u}/4$, $\tau_\m{u}/2$ and $3\tau_{u}/4$ (arrows). Nonclassical correlations break its  spatial symmetry, shifting it toward the wave front that moves with the Lieb-Robinson velocity, therefore boosting the entropy flux. Parameters are $\omega=1$, {$\beta=0$}, $J = 1$ and $N=10$.}
  \label{fig1}
\end{figure*}

\emph{Quantum correlated spin chain.}  We consider a quantum wire consisting of a linear $XY$ chain with $N$ qubits and inhomogeneous nearest-neighbor interactions \cite{chr04}
\begin{align}\label{eq:H}
  H &= \omega \sum\limits_{j=1}^N \sigma^z_j + \sum\limits_{k=1}^{N-1} {J_k}
  (\sigma^x_k\sigma^x_{k+1}+\sigma^y_k\sigma^y_{k+1}),
\end{align}
where $\omega$ is the level spacing  and  $J_k= ({\lambda}/{4})\sqrt{k\cdot(N-k)}$ are mirror-symmetric  interaction constants between adjacent qubits, with coupling strength $\lambda$ (we put $\hbar = 1$). We choose $\lambda = 4J/N$, such that the maximum coupling constant is $\m{max}\{J_k\} = J$, independent of $N$, where $J$ sets the characteristic energy scale \cite{chr05,com}. The first spin $(j=1)$ acts as the sender and the last one $(j=N)$ as the receiver. Information is transmitted along the wire by dynamically transferring the state $\rho^{t=0}_1$ of the first spin to the state $\rho^{t=\tau_\m{u}}_N$ of the last  spin in time $\tau_\m{u}$. The couplings  $J_k$ are engineered such that after time $\tau_\m{u} = \pi/\lambda$, state transfer is perfect, with unit fidelity, $F(\rho^{0}_1,\rho^{\tau_\m{u}}_N) = \text{Tr}\big [{\scriptstyle \sqrt{\sqrt{\rho^{0}_1}\rho^{\tau_\m{u}}_N\sqrt{\rho^{0}_1}}}\big]^2=1$ \cite{chr04}. For instance, if the first spin is initially in the excited state $|1\rangle \langle 1|$ and all the remaining spins are in their  ground state $|0\rangle \langle 0|$, then, after  time $\tau_\m{u}$, all the qubits will be in their ground state, except the last one which will be in the excited state. The time $\tau_\m{u}$ has been shown  to  correspond to the minimal possible evolution time, hence, to the quantum speed limit of perfect state transfer in one dimension \cite{yun06}.

Following Refs.~\cite{leb66,bow67,pen83,bek88,bek90}, we begin by taking the state of the first qubit  to be thermal at inverse temperature $\beta$, $\rho_1^0=\rho^\mathrm{th}_1 = \exp(-\beta\omega  \sigma^z_1)/Z$,  with partition function $Z$ (we shall relax this assumption later on). In this case, the  information current is given by the von Neumann entropy rate (in bits) \cite{leb66,bow67,pen83,bek88,bek90}.  Instantaneous information and energy fluxes at the receiver are accordingly 
\begin{align}
 \!\dot{I} = -\partial_t \Tr[\rho^t_N\log(\rho^t_N)] \,\,\, \text{ and } \,\,\, \dot{E} = \partial_t \Tr[\rho^t_N\omega\sigma^z_N].
\end{align}
 We further prepare the second spin $(j=2)$ in a thermal state at the same inverse temperature, $\rho_2^0=\rho^\mathrm{th}_2 = \exp(-\beta\omega  \sigma^z_2)/Z$ (we shall again relax this assumption in the following), and initialize the  remaining qubits  in their respective ground states. The total initial state is then of the form   $\rho(0) = \rho^0_{12} \otimes \dyad{0}^{\otimes (N-2)}$.

We  examine two   different scenarios: (i) in the first case, the first two qubits are uncorrelated, $\rho^0_{12, \m{u}} = \rho_1^\mathrm{th} \otimes \rho_2^\mathrm{th}$, whereas (ii) in the second case, they are quantum correlated, $ \rho^0_{12,\m{c}} = \rho^\mathrm{th}_1 \otimes \rho^\mathrm{th}_2 + \chi$, with a correlation term $\chi = -i\alpha(\sigma^+_1 \sigma^-_2 - \sigma^-_1 \sigma^+_2)$, with amplitude $\alpha$ and  spin ladder operators $\sigma_j^\pm = \sigma_x \pm i\sigma_y$. This correlated state is such that the respective reduced density operators coincide with the thermal density matrices  of the uncorrelated  state \cite{mic19}; this allows one to directly compare the two situations. The parameter $\alpha$  controls the amount of quantum correlations; it    is upper bounded, $\abs{\alpha} \leq {1}/{(4\cosh^2\beta)} \le {1}/{4}$, to ensure positivity of the density matrix. The state $ \rho^0_{12,\m{c}}$ has nonzero geometric quantum discord defined
as $D_\m{g}(\rho)=\min_{\sigma\in\mathcal{C}}2\|\rho-\sigma\|^{2}$, where $\mathcal{C}$
is the set of all  classically correlated states \cite{mod12,ber17,hu18}. A nonzero value of the geometric discord indicates nonclassical  correlations \cite{dak10,gir12}. We find  that the geometric quantum discord increases quadratically with the amplitude of the correlation term, $D_\m{g}(\rho^0_{12,\m{c}}) = 2\alpha^2$.

The presence of initial nonclassical correlations  strongly affects  the propagation of information and energy  in the medium.
\Cref{fig1} examines  the impact of local quantum correlations  on information transmission along the $XY$ spin chain \eqref{eq:H}. \Cref{fig1}a displays the fidelity 
$F(\rho^{0}_1,\rho^t_N)$ between emitter and receiver as a function of time   for $N=10$ qubits and $\alpha = 1/4$. While the first spin state is exactly transmitted to the final qubit in time $\tau_\m{u}$ for uncorrelated initial states (blue), perfect state transfer is achieved much faster, after time $\tau_\m{c}$, for correlated initial conditions (red). The time $\tau_\m{u}$, therefore, no longer corresponds to the ultimate quantum speed limit, as for the uncorrelated state \cite{yun06}. Interestingly, faithful transfer occurs a second time at time $\tau_\m{u}$.  This enhanced transmission is accompanied by an increase of  the maximum information flux  ${\dot{I}}_\text{max}$, as well as of the maximum energy current $ \dot{E}_\text{max}$ (arrows in \cref{fig1}b). We will see below that  those boosts are directly related to the amount of quantum correlations in the initial state. We also observe that the  information-energy flow inequality, $\dot I^2 \leq \pi  \dot E/3$,  is obeyed in both instances for $t\leq\tau_\m{u}$, respectively $t\leq\tau_\m{c}$ (\cref{fig1}b). After full information transfer, the roles of sender and receiver are switched, and the currents are reversed.
We mention  that the fidelity, as well  as the information and energy fluxes can be evaluated analytically for the spin chain \eqref{eq:H}  with the help of the Jordan-Wigner transformation (Supplemental Material).

\Cref{fig1} further shows the temporal evolution of the information $I$ and of the energy $E$ of the receiving qubit $\rho_N$. In the uncorrelated case, the two quantities reach the initial values of the sender, $I= \ln 2$ and $E = 0$, simultaneously at time $\tau_\m{u}$. By contrast, information and energy display a different time dependence in the presence of correlations. The von Neumann entropy behaves similarly to the fidelity $F(\rho^{0}_1,\rho^t_N)$, attaining  its initial, and maximum, value at time $\tau_\m{c}$. However, the energy  overshoots the initial energy of the sender, reaching its maximum after $\tau_\m{c}$, until it gets reflected at the end of the chain and flows back  such that it coincides with the initial energy of the sender at time $\tau_\m{u}$, like in the uncorrelated case. This remarkable behavior is an immediate consequence of the initial quantum correlations.

\emph{Enhanced information transmission}.  In order to quantitatively analyze the boosted transfer of information along the  quantum wire, we next use the out-of-time-order correlator (OTOC) \cite{swi18,lew19,gar23}. We specifically  consider the expectation value of the squared commutator of two local, spatially separated (Hermitian) operators $W(x,t)$ and $V(y,0)$, $C(x,y,t) = {\langle [W(x,t),V(y,0]^2\rangle} = 2- 2 \m{Re}[F(x,y,t)]$  for unitary dynamics \cite{lar69,swi18,lew19,gar23}. The out-of-time-order correlator, $  F(x,y,t) = \langle W(x,t)V(y,0)W(x,t)V(y,0)\rangle$, characterizes the spread of quantum information along the spin chain, commonly referred to as information scrambling \cite{swi18,lew19,gar23}. For concreteness, we choose the two single-site Pauli operators $W(x,t) = \sigma^z_x(t)$ and $V(y=3,0) = \sigma^z_3(0)$; the time-evolved Heisenberg operator $W(x,t)$ can thus be regarded as a local probe (at site $x$ and time $t$) of the information emitted by the (correlated) sender at $t=0$.

\Cref{fig1}d compares the stroboscopic, wave-like evolution of the correlation function $C(x,3,t)$ at three different times, $\tau_\m{u}/4$, $\tau_\m{u}/2$ and $3\tau_\m{u}/4$, for uncorrelated (blue)  and correlated (red) initial states;  we have evaluated $C(x,3,t)$ numerically for a chain of $N=250$ spins \cite{lui17,lin18,bao20}. The correlator exhibits three peaks that travel from left to right with, in both instances, the {same} wavefront which moves close to the Lieb-Robinson velocity $v_\m{LR} = 2J$.  The latter velocity provides a fundamental upper bound to the spread of information in locally interacting quantum systems \cite{lie72}. Information can hence not propagate faster than $v_\m{LR}$. However, we note that the correlator is no longer {spatially symmetric} in the presence of quantum correlations (red). The weight of the distribution $C(x,3,t)$ is indeed shifted towards the wave front, with {higher peaks near the wave front} compared to the case without   initial correlations. For example, at $t = \tau_\m{u}/4$, quantum correlations increase the mean $\overline x = \int dx\, x \,C(x,3, \tau_\m{u}/4) /\int dx\,  C(x,3, \tau_\m{u}/4)$ by $8.15\%$ from  37.41 to 40.46 for the parameters of \cref{fig1}d. As a consequence, more information flows per unit time towards the receiving end of the chain, giving rise to the faster state transfer seen in \cref{fig1}abc. Symmetries of the out-of-time-order correlator are further elaborated on in the Supplemental Material.

\begin{figure*}[t]
    \begin{tikzpicture}
      \node (a) [label={[label distance=-0.2 cm]143: \textbf{(a)}}] at (0,0) {\includegraphics[width=8.9cm]{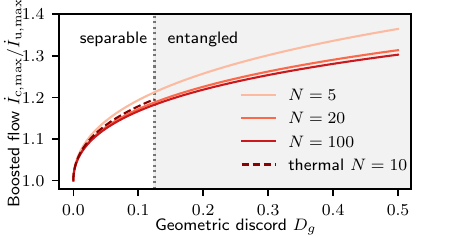}}; 
      \node (b) [label={[label distance=-0.2 cm]143: \textbf{(b)}}] at (8,0) {\includegraphics[width=8.9cm]{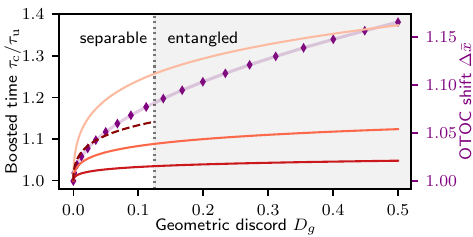}};	
    \end{tikzpicture}
  \caption{Boosted information flow and transfer time. a) Maximum information current ratio, $\dot I_\text{c,max}/ \dot I_\text{u,max}$, and b) transfer time ratio, $\tau_\m{u}/\tau_\m{c}$, as a function of the geometric quantum discord $D_\m{g}$ of  the $X$ state, Eq.~\eqref{eq:X}, for various values of the chain length $N$ (solid lines). Both quantities increase monotonically with the amount of quantum correlations, which does not need to be entangled. The dashed line represents the correlated thermal state of Fig.~1.  Purple diamonds display the shift toward the wave front, $\Delta \bar x = \bar x_\m{c} - \bar x_\m{u}$, of the averaged out-of-time-order correlator $C(x,3,\tau_\m{u}/4)$  induced by the nonclassical correlations. {Parameters for the $X$ state are $a=d=w=0$, $b=c=1/2$ and $z = i\alpha$}.}
  \label{fig2}
\end{figure*}

Let us now investigate the role of  initial nonclassical correlations in more detail. \Cref{fig2} clearly indicates that both the amplified maximum information flow, $\dot I_\text{c,max}/ \dot I_\text{u,max}$ (\cref{fig2}a) and the  enhanced transfer time, $\tau_\m{u}/\tau_\m{c}$ (\cref{fig2}b), monotonically increase with the amount of quantum correlations, quantified by the geometric discord of the initial state $D_\m{g}(\rho^0_{12,\m{c}})$ (dark red, dashed). We may hence conclude that nonclassical correlations are a local physical resource that can boost communication rates in a wire. Interestingly, the state $\rho^0_{12,\m{c}}$ is separable with zero concurrence \cite{woo98} for all values of $\alpha$. We observe, in fact, no information transmission boost when a maximally entangled Bell state is used (Supplemental Material). This quantum advantage  is therefore closer  to the deterministic quantum computation with one qubit (DQC1) model \cite{kni98}, where the quantum resource for computational speedup is associated with discord, and not with entanglement \cite{dat08,lan08,wan19}. 

\emph{Criterion for enhanced transmission}. 
The above effect is not restricted to correlated thermal states or to $XY$ spin chains. We next consider a quantum spin chain of the form $ H = \omega \sum_j \sigma^z_j + H_\text{int}$ with general coupling Hamiltonian $H_\text{int}$, and  an initial correlated state $ \rho^0_{12,\m{c}} = {\rho_\m{u} + \chi}$ with diagonal part $\rho_\m{u}$. We furthermore separate the unitary time evolution into  uncorrelated and correlated parts,  $ \rho_\m{c}(t) = \rho_\m{u}(t) + U \big(\chi \otimes \dyad{0}^{\otimes (N-2)}\big) U^\dagger$ with $U=\exp(-iHt)$. In order to boost state transfer, we demand that, at short times, more probability is shifted towards the state $\ket{\psi_2} = \ket{01}\otimes \ket{0}^{\otimes (N-2)}$ than in the uncorrelated case, that is, $  \bra{\psi_2} \dot{\rho}_\m{c}(\dd{t})\ket{\psi_2}> \bra{\psi_2}\dot{\rho}_\m{u}(\dd{t})\ket{\psi_2} $. Since the local initial states are equal in both cases, this condition implies  that $\bra{01}[H_\m{int},\chi]\ket{01} > 0$ (Supplemental Material). Note that only the interaction Hamiltonian matters because the diagonal part does not connect different sites of the chain. A necessary requirement for enhanced transmission is therefore
\begin{align}\label{eq:condition}
  \left[H_\m{int}, \chi\right] \neq 0.
\end{align}
Boosted entropy flux is thus {generally} possible whenever the correlation term $\chi$ does not commute with the interaction Hamiltonian $H_\text{int}$. The latter criterion underscores the quantum origin of the phenomenon.

As an illustration, we consider the quantum $XY$ chain \eqref{eq:H} with a correlated state given by an  $X$ state \cite{yu07,ali10,che11,que12}
\begin{align}
  \rho^0_{12,\m{c}}= \rho_X = 
  \begin{pmatrix}
    a & 0 & 0 & w \\
    0 & b & z & 0 \\
    0 & z^\ast & c & 0 \\
    w^\ast & 0 & 0 & d 
  \end{pmatrix},
  \label{eq:X}
\end{align}
with $|z| \le \sqrt{bc}$ and $|w| \le \sqrt{ad}$. Such X states define a general class of quantum correlated bipartite qubit states that, for instance,  include Werner states and Bell states \cite{yu07,ali10,che11,que12}. The corresponding reduced state of the first spin to be transferred is $\rho^0_1 = \Tr_2[\rho_X] =\text{diag}(a+b,c+d)$. Condition \eqref{eq:condition} requires that $|z| \neq 0$, while the entries ($w,w^\ast,a,d$), that belong to the commuting part, may be chosen arbitrarily. The concurrence  of the $X$ states is given by $C(\rho_X) = 2\, \m{max}\{0,z-\sqrt{ad},w-\sqrt{bc}\}$ \cite{yu07,ali10,che11,que12}, whereas their geometric discord reads $D_\m{g}(\rho_X) 
= \min\{4(w^2+z^2), (a-c)^2+(b-d)^2+2(w+z)^2\}/2$ \cite{yu07,ali10,che11,que12}. $X$ states are hence entangled if and only if $bc < |w|^2$ or $ad < |z|^2$, but both conditions cannot be satisfied simultaneously  \cite{ali10}. States with $w = w^\ast = 0$ and $ad = |z|^2 = \alpha^2$, with $\alpha < 1/4$ (like the thermal correlated states considered before) are thus not entangled. However, in general $D_\m{g}(\rho_X)= 2\alpha^2$ for $z = i\alpha$, $\alpha\le 1/2$, revealing the presence of nonclassical correlations. 

\Cref{fig2} presents  the amplified maximum information flow, $\dot I_\text{c,max}/ \dot I_\text{u,max}$ (\cref{fig2}a), and the enhanced transfer time, $\tau_\m{u}/\tau_\m{c}$ (\cref{fig2}b), for the X state \eqref{eq:X} (solid lines) as a function of the geometric quantum discord $D_\m{g}(\rho_X)$, for $N= 5, 20$  and $100$ spins, when $\alpha$ is varied. As for the thermal correlated state, $ \rho^0_{12,\m{c}} = \rho^\mathrm{th}_1 \otimes \rho^\mathrm{th}_2 + \chi$, both quantities grow monotonically with the amount of nonclassical correlations. However, since $D_\m{g}(\rho_X)$ can take larger values when the $X$ state is entangled, the quantum enhancement is more pronounced. For example, for  the $N=11$ sites of the state transfer experiment reported in Ref.~\cite{cha16}, a maximal geometric discord of 1/2 would increase the maximal information flux by $32.6\%$ and decrease the transfer time by $18.7\%$. \Cref{fig2}b additionally shows that these boosts are accompanied by an increased shift toward the wavefront, $\Delta \bar x = \bar x_\m{c} - \bar x_\m{u}$, of the averaged out-of-time-order correlator $C(x,3,\tau_\m{u}/4)$ (purple diamonds), confirming that this effect finds its origin
in the breaking of the spatial symmetry of the out-of-time-order correlator by the quantum correlations. The increased maximum entropy current and transfer time both exhibit a square-root dependence on the geometric quantum discord $D_\m{g}$, that is, a linear dependence of the correlation strength $\alpha$. We additionally note that due to the singular derivative of the square root at the origin, only a small amount of quantum discord is required to achieve a significant transmission improvement.

\textit{Conclusions.} Entanglement is a fundamental resource for quantum applications that outperform their classical counterparts \cite{gis07}.  However, entangled states are a costly and fragile resource that is difficult to prepare and maintain over long distances, owing to the detrimental effect of decoherence \cite{sch07}. We have  shown that information propagation can be enhanced in a quantum correlated medium without requiring entanglement distribution between emitter and receiver. Remarkably,   local nonclassical correlations with nonzero discord  between only the first two qubits at the emitter suffice to boost the information flow for spin chains of arbitrary length. While information cannot propagate faster than the Lieb-Robinson velocity, the breaking of the spatial symmetry of the out-of-time-order correlator induced by the quantum correlations leads to an augmented entropy current. Our results indicate that quantum correlated media are a useful resource for quantum enhanced communication.

\textit{Acknowledgements.} 
We acknowledge  financial support from the Vector Foundation and from the German Science Foundation (DFG) under project FOR 2724.

\clearpage
\widetext
\begin{center}
\textbf{\large Supplemental Material: Boosting information transfer in a quantum correlated medium}
\end{center}
\setcounter{equation}{0}
\setcounter{figure}{0}
\setcounter{table}{0} 
\setcounter{page}{1}
\setcounter{secnumdepth}{4}
\makeatletter
\renewcommand{\theequation}{S\arabic{equation}}
\renewcommand{\thefigure}{S\arabic{figure}}
\renewcommand{\bibnumfmt}[1]{[S#1]}
\renewcommand{\citenumfont}[1]{S#1}


\renewcommand{\figurename}{Supplementary Figure}
\renewcommand{\theequation}{S\arabic{equation}}
\renewcommand{\thefigure}{S\arabic{figure}}
\renewcommand{\bibnumfmt}[1]{[S#1]}
\renewcommand{\citenumfont}[1]{S#1}

In this Supplemental Material, we will (I) present the analytical evaluation of  the time evolution of the local populations using the Jordan-Wigner transformation, (II) derive the condition for faster state transfer, (III) analyze the out-of-time-order correlator of the perfect state transfer and the uniformly coupled quantum spin chain and (IV) show that there is no information boost for Bell states.

\section{Analytical results}
\subsection{Jordan-Wigner transformation}
In this section we compute the local populations of the spin chain and from them derive the expressions for the state transfer fidelity and the von Neumann entropy.
Our starting point is the perfect state transfer Hamiltonian for a chain of $N$ qubits with nearest neighbor interaction, Eq.~(2) of the main text, 
\begin{align} 
    H = \omega \sum_{j=1}^N \sigma^z_j + \sum_{k=1}^{N-1} {J_k}
    (\sigma^x_k\sigma^x_{k+1}+\sigma^y_k\sigma^y_{k+1}) \quad
    \qq{with} J_k &= \frac{\lambda}{4}\sqrt{k\cdot(N-k)},
    \label{seq:H}
\end{align}
where $\omega$ is the level spacing of the qubits and the $J_k$ are the interaction constants between adjacent qubits.
Let $\rho(0)$ be the composite initial state of the full qubit chain.
The interactions $J_k$ are engineered in such a manner that after a constant time $\tau_\m{u} = \pi/\lambda$ (which we consider the final time of the process), the state of the chain corresponds to the
mirror image of its initial configuration (the initial local states of the individual qubits are swapped with their mirror image states) \cite{Schr04,Schr05}.
To avoid unphysical Hamiltonians and violation of causality, we normalize the interaction constants such that their maximum value is bounded.
Specifically, we set $\lambda = 4J/N$, so that $\m{max}\{J_k\} = J$, independent of $N$ \cite{Schr05}. The perfect transfer time then scales linearly with the system size $N$, and we obtain $\tau_\m{u} = N\pi/(4J)$.

The initial states considered in the main text can be written in the form 
\begin{align}
    \rho(0) = \rho^0_{12} \otimes \dyad{0}^{\otimes (N-2)},
\end{align}
where
\begin{align}
    \rho^0_{12} = \rho_X = 
    \begin{pmatrix}
      a & 0 & 0 & w \\
      0 & b & z & 0 \\
      0 & z^\ast & c & 0 \\
      w^\ast & 0 & 0 & d 
    \end{pmatrix},
    \label{seq:X}
\end{align}
is an $X$ state \cite{Syu07,Sali10,Sche11,Sque12}. 
Concretely, we consider initial states that can be parametrized as  
\begin{align}
    \rho^0_{12} = \rho_1 \otimes \rho_2 + \chi,
    \label{eq:thermal}
\end{align}
with 
\begin{align}
    \rho_j = 
    \begin{pmatrix}
        p_j & 0\\
        0 & 1-p_j
    \end{pmatrix}
    \quad \text{and} \quad
    \chi = -i\alpha(\sigma^+_1 \sigma^-_2 - \sigma^-_1 \sigma^+_2),
    \label{eq:JW-initial}
\end{align}
with real parameter $\alpha \in [0,1/4]$.
For any $\alpha \neq 0$, these states have non-zero geometric quantum discord \cite{Syu07,Sali10,Sche11,Sque12}
\begin{align}
    D(\rho_X) = \frac{1}{2}\min\{4(w^2+z^2),(a-c)^2+(b-d)^2+2(w+z)^2\},
\end{align}
but are not necessarily entangled, since their concurrence is given by \cite{Syu07,Sali10,Sche11,Sque12}
\begin{align}
    C(\rho_X) = 2 \m{max}\{0,z-\sqrt{ad},w-\sqrt{bc}\}.
\end{align}
For initial states of the form \cref{eq:JW-initial}, the evolution of the local populations can be conveniently calculated by performing a Jordan-Wigner transformation \cite{Stak99}.  
The transformed Hamiltonian reads
\begin{align}
    H_\m{JW} = \omega\sum_{k=1}^N (2 c^\dagger_k c_k - \mathds{1}) + 2\sum_{k=1}^{N-1} J_k(c^\dagger_k c_{k+1} + c^\dagger_{k+1} c_k), 
\end{align}
where $c^\dagger_k$ and $c_k$ are the respective fermionic ladder operators. They are related to the usual Pauli operators by
\begin{align}
    c_k = \exp(i\pi\sum^{k-1}_{n=1} c^\dagger_nc_n) \sigma^-_k, \qquad c_j^{\dagger} = \exp(i\pi\sum^{k-1}_{n=1} c^\dagger_nc_n) \sigma^+_k.
\end{align}
If we restrict ourselves only to the evolution of the local populations and nearest-neighbor correlations, we obtain an effective von Neumann equation
\begin{align}
    \dot{Z}(t) = i[\Omega,Z(t)],
    \label{eq:effective}
\end{align}
for the correlations $Z_{jk}(t) = \langle\sigma^+_j\sigma^-_k\rangle(t)$, representing the populations on the diagonal $Z_{jj}(t) = p_j(t)$, and nearest-neighbor correlations $Z_{j,j\pm1} =\langle\sigma^+_j\sigma^-_{j\pm1}\rangle(t)$ on the off-diagonals.
The matrix $\Omega = \m{diag}(2\vb*{J},\omega,2\vb*{J})$, where $\vb*{J}=(J_1,J_2\ldots,J_{N-1})^\mathrm{T}$, acts as an effective tridiagonal Hamiltonian with the interaction strengths on the off-diagonals and the qubit level spacing $\omega$ on the diagonal.

\subsection{One-excitation subspace}
The class of initial states can be further extended beyond \cref{eq:thermal} by 
considering initial states that lie in the single excitation sector, i.e. in the subspace that contains only one spin excitation.
Since the total magnetization in the $z$-direction is conserved,  $[H,\sum_k \sigma^z_k] = 0$, the Hamiltonian can be decomposed into mutually orthogonal subspaces corresponding to a fixed number of excitations and the evolution inside each subspace is consequently independent from the remaining Hilbert space.
If we denote by 
\begin{align}
    P_1 = \sum_k \dyad{1_k}
\end{align}
the projector on the single excitation sector with 
\begin{align}
    \ket{1_k} \in \{\ket{10\cdots 0}, \ket{010\cdots 0}, \ket{0010\cdots 0}, \ldots, \ket{0\cdots 01}\},
\end{align}
then the projection of the initial $X$ state \cref{seq:X} becomes 
\begin{align}
    P_1 \rho_X P_1 = 
    \begin{pmatrix}
        b & z\\
        z^\ast & c.
    \end{pmatrix}
    \label{eq:x-sector}
\end{align}
We choose $z = i\alpha$, where $\alpha \in [0,1/2]$ can now be twice as large as before.
The one excitation sector thus supports only $X$ states for which $a = d = w = 0$.

The Hamiltonian inside this subspace becomes
\begin{align}
    P_1 H P_1 = \Omega,
\end{align}
which is equal to the adjacency matrix of the physical graph corresponding to the spin chain and it coincides with the effective Hamiltonian of the Jordan-Wigner transform, \cref{eq:effective}.
The local populations on the full Hilbert space can be written as
\begin{align}
    p_j(t) = \Tr[\rho(t)\Pi_j], \quad \text{with projectors} \quad
    \Pi_j = 
    \mathds{1}^{\otimes j-1} \otimes 
    \begin{pmatrix}
        1 & 0\\
        0 & 0
    \end{pmatrix}
    \otimes
    \mathds{1}^{\otimes N-j}.
\end{align}
In the single excitation sector, the projector $\Pi_j$ reduces to 
\begin{align}
    P_1 \Pi_j P_1
    = \dyad{j},
\end{align}
such that the populations coincide with the ones obtained from the Jordan-Wigner transform
\begin{align}
    p_j(t) = Z_{jj}(t),
    \label{eq:magnetization}
\end{align}
following from the same effective von Neumann equation \eqref{eq:effective}.
Both approaches, the Jordan-Wigner transformation and projecting onto the single excitation sector, thus lead to the same evolution equation for the local populations differing only in the permitted values for $\alpha$.

\subsection{Local populations}
For initial states that either lie in the single excitation subspace, \cref{eq:x-sector}, or can be parametrized as \cref{eq:thermal}, we accordingly only need to consider the effective von Neumann equation \cref{eq:effective}. The solution reads 
\begin{align}
    Z(t) = {e^{i\Omega t}}Z(0)e^{-i\Omega t},
\end{align}
where the time evolution operator \cite{Schr04}
\begin{align}
    U(t) = {e^{i\Omega t}} = \exp(i\lambda S_x t)
    \label{eq:unitary}
\end{align}
represents the rotation of a spin $(N-1)/2$ particle around the $x$-axis.

In this description, the initial states take the form 
\begin{align}
    Z(0) =
    \begin{bmatrix}
        p_1(0) & i\alpha & \cdots & 0\\
        -i\alpha & p_2(0) & \cdots & 0\\
        \vdots & \vdots & \ddots & \vdots\\
        0 & 0 & \cdots & 0
    \end{bmatrix}
    \label{eq:initial-state}
\end{align}
corresponding to the initial $X$ state in Eq.~(14) of the main text.
The populations then follow from the diagonal elements of $Z(t)$, which can be straightforwardly computed in terms of the matrix elements of the unitary \eqref{eq:unitary}
\begin{align}
    p_k(t) = 
    Z_{kk}(t) = 
    p_1 |U_{k1}|^2 + p_2 |U_{k2}|^2 - 2 \alpha\m{Im}[U_{k1}^\ast U_{k2}].
\end{align}
We are in particular interested in the population of the last spin
\begin{align}
    p_N(t) = 
    Z_{NN}(t) = 
    p_1 |U_{N1}|^2 + p_2 |U_{N2}|^2 - 2 \alpha\m{Im}[U_{N1}^\ast U_{N2}].
\end{align}
The matrix elements $U_{kl}$ follow from the Wigner $d$-matrix \cite{Swig12} 
\begin{align}
\label{23}
    d^j_{m^\prime m}(\theta) = [(j+m^\prime )!(j-m^\prime )!(j+m)!(j-m)!]^{1/2} \sum_{s=s_{\mathrm{min}}}^{s_{\mathrm{max}}} \left[\frac{(-1)^si^{m-m^\prime} \cos(\frac{\theta}{2})^{2j+m-m^\prime -2s}\sin(\frac{\theta}{2})^{m^\prime -m+2s}}{(j+m-s)!s!(m^\prime -m+s)!(j-m^\prime -s)!} \right],
\end{align}
where $s_\m{min} = \m{max}\{0,m-m^\prime\}$ and $s_\m{max} = \m{min}\{j+m,j-m^\prime\}$ and $\theta = \lambda t$.
\Cref{eq:unitary} represents the elements of the rotation matrix of a spin $j = (N-1)/2$ particle with magnetizations running over $m = -j,-j+1,\ldots,j-1,j$.
We therefore have 
\begin{align}
    U_{kl} 
    &= d^{(N-1)/2}_{\frac{-(N-1)}{2}+k-1,\frac{-(N-1)}{2}+l-1}(\theta),\\
    &= [(k-1)!(N-k)!(l-1)!(N-l)!]^{1/2} \sum_{s=s_{\mathrm{min}}}^{s_{\mathrm{max}}} \left[\frac{(-1)^si^{l-k} \cos(\frac{\theta}{2})^{(N-1)+l-k-2s}\sin(\frac{\theta}{2})^{k-l+2s}}{(l-1-s)!s!(k-l+s)!(N-k-s)!} \right],
\end{align}
with $s_\m{min} = \m{max}\{0,l-k\}$ and $s_\m{max} = \m{min}\{l-1,N-k\}$.
Plugging in $(k,l) = (1,N)$ and $(k,l) = (2,N)$, we get for the individual contributions
\begin{align}
    [(k-1)!(N-k)!(l-1)!(N-l)!]^{1/2}
    = 
    \begin{dcases}
        (N-1)!, & (k,l) = (1,N)\\
        \sqrt{(N-1)!(N-2)!}, & (k,l) = (2,N)
    \end{dcases}    
    \label{eq:t1}
\end{align}
\begin{align}
    s_\m{min} &= \m{max}\{0,l-k\} = 
    \begin{dcases}
    N-1, & (k,l) = (1,N)\\
    N-2, & (k,l) = (2,N)
    \end{dcases}
    \label{eq:t2}\\
    s_\m{max} &= \m{min}\{l-1,N-k\} = 
    \begin{dcases}
        N-1, & (k,l) = (1,N)\\
        N-2, & (k,l) = (2,N)
    \end{dcases}
    \label{eq:t3}
\end{align}
\begin{align}
    (-1)^si^{l-k} = 
    \begin{dcases}
        (-1)^{N-1}i^{N-1}, & (k,l) = (1,N)\\
        (-1)^{N-2}i^{N-2}, & (k,l) = (2,N)
    \end{dcases}
    \label{eq:t4}
\end{align}
\begin{align}
    (l-1-s)!s!(k-l+s)!(N-k-s)! =
    \begin{dcases}
    (N-1)!, & (k,l) = (1,N)\\
    (N-2)!, & (k,l) = (2,N)
    \end{dcases}
    \label{eq:t5}
\end{align}
\begin{align}
    (N-1)+l-k-2s &= 
    \begin{dcases}
    0, & (k,l) = (1,N)\\
    1, & (k,l) = (2,N)
    \end{dcases},
    \label{eq:t6}\\ 
    k-l+2s &=
    \begin{dcases}
        N-1, & (k,l) = (1,N)\\
        N-2, & (k,l) = (2,N)
    \end{dcases}.
    \label{eq:t7}
\end{align}
Finally, using \cref{eq:t1,eq:t2,eq:t3,eq:t4,eq:t5,eq:t6,eq:t7} together with \cref{eq:pn}, we obtain the time evolution of the local population of the $N$th qubit of the chain as
\begin{align}
    p_N(t) = p_1(0) \left[\sin(\frac{\theta}{2})\right]^{2(N-1)} 
    + p_2(0) (N-1) \left[\sin(\frac{\theta}{2})\right]^{2(N-2)}\cos(\frac{\theta}{2})^2 
    + 2\sqrt{N-1}\alpha \left[\sin(\frac{\theta}{2})\right]^{2N-3}\cos(\frac{\theta}{2}).
    \label{eq:pn}
\end{align}
The last term in the above equation is always greater or equal to zero for $\theta \in [0,\pi]$ which is equivalent to $t \in [0,\tau_\m{u}]$, where $\tau_\m{u} = \pi/\lambda = N\pi/(4J)$ is the transfer time of the uncorrelated chain.

\subsection{Fidelity and von Neumann entropy}
The Hamiltonian \eqref{eq:H} does not create coherences.
Therefore, the reduced density matrices of the local qubits are always diagonal.
Knowledge of the local populations is hence sufficient to compute the fidelity and the von Neumann entropy as will be illustrated below.
For two single-qubit states $\rho$ and $\sigma$, the fidelity can be computed according to \cite{Sbar13}
\begin{align}
    F(\rho,\sigma) 
    = O(\rho,\sigma) + \sqrt{[1-O(\rho,\rho)][1-O(\sigma,\sigma)]},
    \label{eq:fidelity}
\end{align}
where $O(\rho,\sigma) = \Tr[\rho\sigma]$.
To detect the arrival of a state at the end of the chain, we compute the fidelity of the initial state at the first site with the local state at the last site.
In the absence of coherences, the fidelity reduces to
\begin{align}
    F(\rho_1(0),\rho_N(t)) = \left[\sqrt{(1-p_1(0))(1-p_N(t))}+\sqrt{p_1(0)p_N(t)}\right]^2.
\end{align}
The initial local state is hence transferred with unit fidelity when $p_N(t) = p_1(0)$.
Initially, we have $p_N(0) = 0$ (the last spin starts in the ground state).
In the absence of quantum correlations, for $t>0$, there will then be a monotonic increase in the population until the state is transferred (cf. \cref{eq:pn}).
For every positive value of $\alpha$ the last term in \cref{eq:pn} is always positive during the transfer time, any amount of nonzero quantum discord can lead to faster state transfer.
Note that even for $\alpha < 0$, the Hamiltonian \cref{seq:H} still guarantees a transfer time of at most $\tau_\m{u}=\pi/\lambda$ where the correlation term in \cref{eq:pn} vanishes. 

In order to evaluate the information received at the end of the chain, we use the von Neumann entropy.
Because local states always remain diagonal, the von Neumann entropy is simply given by 
\begin{align}
    I 
    = -\Tr[\rho_N(t)\log(\rho_N(t))]
    = -p_N(t)\log(p_N(t)) - (1-p_N(t))\log(1-p_N(t)).
\end{align}
The information flow can accordingly be conveniently expressed as 
\begin{align}
    \dot{I} = - \dot{p}_N(t)\log(\frac{p_N(t)}{1-p_N(t)}).
\end{align}
where $p_N(t)$ is given in \cref{eq:pn}.
\newpage

\section{Condition for faster state transfer}
In quantum systems with local interactions like the perfect state transfer model \cref{seq:H}, every order of the power series of the time evolution operator $U=\exp(-iHt)$ can only connect adjacent sites. For instance, the evolution of a state $\ket{\psi_1} = \ket{1}\otimes \ket{0}^{\otimes (N-1)}$ is given by
\begin{align}
    U \ket{\psi_1} 
    = \exp(-iHt)\ket{\psi_1} 
    = \sum_{n=0}^\infty \frac{(-i)^n}{n!} t^n H^n\ket{1}\otimes \ket{0}^{\otimes (N-1)} .
    \label{eq:moments}
\end{align}
Every order of the Hamiltonian can shift the excitation through the chain by at most one site such that after applying the Hamiltonian $n$ times, the state is generally in a superposition of the form
\begin{align}
    H^n \ket{1}\otimes\ket{0}^{N-1} = \sum_{s=1}^n \sum_{j=1}^{M(n)} c_j(n,s) \ket{s_{\sigma(j)}} \otimes \ket{0}^{N-1-n}
\end{align}
where $s \le n$ corresponds to a fixed number of excitations and $\ket{s_{\sigma(j)}}$ denotes all possible permutations of ones and zeros for a given $s$ (all the possibilities to distribute $s$ excitations over $n$ sites) with 
\begin{align}
    M(n) = 
    \begin{pmatrix}
    n \\ s    
    \end{pmatrix}
    = \frac{n!}{s!(n-s)!}
\end{align}
elements.
The $c_j(n,s)$ are arbitrary coefficients that depend on the specifics of the Hamiltonian.

Every order of $H$ is accompanied by a corresponding order of $t$ such that, at each step in time, any excitation can at most reach its immediate neighbor. 
Information propagation is moreover upper bounded by the Lieb-Robinson velocity $v_\m{LR} = 2J$ and can thus spread at most linearly in time. This limitation cannot be overcome even in the presence of correlations.
However, transmission of a qubit end-state along the spin chain is still significantly improved in the presence of nonclassical correlations.
To show this, we consider a quantum spin chain of the general form 
\begin{align}
    H = \sum_j \omega_j\sigma^z_j + H_\text{int}
\end{align}
with possibly different level spacings $\omega_j$ and general nearest-neighbor coupling Hamiltonian $H_\text{int}$.
We assume that the initial correlated state $\rho^0_{12,\m{c}} = \rho_\m{u} + \chi$ can be decomposed into a diagonal part $\rho_\m{u}$ with $\Tr[\rho_\m{u}] = 1$ and correlations $\chi$ with $\Tr[\chi] = 0$. 
We further assume that the local initial states are diagonal and that the Hamiltonian does not generate coherences during the evolution, at least not in the subspace the evolution is restricted to.
We separate the unitary time evolution into uncorrelated and correlated parts
\begin{align}
\rho_\m{c}(t) = \rho_\m{u}(t) + U \chi U^\dagger.
\end{align}
In order to boost state transfer, only the local populations matter and we demand that, at short times, more probability is shifted towards the state $\ket{\psi_2} = \ket{01}\otimes \ket{0}^{\otimes (N-2)}$ than in the uncorrelated case, that is, 
\begin{align}
    \bra{\psi_2} \dot{\rho}_\m{c}(\dd{t})\ket{\psi_2} > \bra{\psi_2}\dot{\rho}_\m{u}(\dd{t})\ket{\psi_2},
    \label{eq:cond1}
\end{align} 
where 
\begin{align}
    \bra{\psi_2} \dot{\rho}_\m{c}(\dd{t})\ket{\psi_2}
    = \bra{\psi_2} \rho_c(0)\ket{\psi_2}
    -i \bra{\psi_2} [H,\rho_\m{u}(0)]\ket{\psi_2}
    -i \bra{\psi_2} [H,\chi]\ket{\psi_2}.
\end{align}
Since the local initial states are equal and the Hamiltonian appears only in first order (thus connecting only the first and second site), condition \eqref{eq:cond1} implies  that $-i\bra{01}[H_\m{int},\chi]\ket{01} > 0$. 
Note that only the interaction Hamiltonian matters because the diagonal part does not connect different sites of the chain. A minimal requirement for enhanced transmission is therefore that the commutator
\begin{align}
    \label{seq:condition}
  \left[H_\m{int}, \chi\right] \neq 0
\end{align}
does not vanish.

In the perfect state transfer model \eqref{eq:H} condition \cref{seq:condition} ensures that local initial correlations on the sender's side, simultaneously reduce the arrival time and increase the maximum information flow $\dot{I}_\m{max}$ (see Figs.~ 1 and 2 of the main text) for, in principle, arbitrary system size $N$ (see \cref{eq:pn}).

\section{Out-of-time-order correlator}
In this section, we provide additional information on the out-of-time-order correlator in the spin chain \eqref{eq:H}.
We numerically compute the squared commutator
\begin{align}
    C(x,y,t) = \langle [W(x,t),V(y,0)]^\dagger[W(x,t),V(y,0)]\rangle,
\end{align}
of the time evolved Heisenberg operator $W(x,t)$ localized at $x$ and the operator $V(y,0)$ localized at $y$ at the initial time.
For operators that are both unitary and Hermitian, the correlator conveniently reduces to
\begin{align}
    C(x,y,t) = 2-2\m{Re}[F(x,y,t)],
    \label{eq:C}
\end{align}
where
\begin{align}
    F(x,y,t) = \langle \sigma^z(x,t)\sigma^z(y,0)\sigma^z(x,t)\sigma^z(y,0)\rangle
    \label{eq:F}
\end{align}
is the out-of-time-order correlator. 
The action of $F(x,y,t)$ can be decomposed into consecutive steps:
the local Pauli operator at site $y$ acts on the initial state $\ket{\psi(y,0)} = \sigma^z(y,0) \ket{\psi}$.
Then, the time evolved operator located at another site $x$ acts on the resulting state yielding $\ket{\psi(y,0;x,t)} = \sigma^z(x,t)\ket{\psi(y,0)}$.
The out-of-time-order correlator  evaluates the overlap, $F(x,y,t) = \braket{\psi(x,t;y,0)}{\psi(y,0;x,t)}$, between the perturbed state $\ket{\psi(y,0;x,t)}$ and a state $\ket{\psi(x,t;y,0)}$, where both actions are performed in reverse order. 
Initially, all Heisenberg operators commute and the overlap is one, yielding a zero correlator $C(x,y,t) = 0$.
As time increases, the time evolution will gradually connect Heisenberg operators at increasingly distant sites.
As soon as two sites $j$ and $k$ have overlapping support, the out-of-time-order correlator will no longer be one, and the correlator $C(x,y,t)\neq 0$ will be different from zero. The latter therefore provides a measure of operator spreading, and is related to information propagation \cite{Sxu24}.

For locally interacting quantum systems (where interactions decay at least exponentially as a function of the distance between sites), the above statement is true in general.
The operator growth rate is bounded from above by the Lieb-Robinson velocity $v_\m{LR}$ \cite{Slie72}.
Generically, the correlator $C(x,y,t)$ spreads through space with a wave front, whose tail has a universal scaling behavior \cite{Skhe18,Sxu20,Sbao20} 
\begin{align}
    C(r,t) \sim \exp(\frac{\Lambda (t-r(t)/v)^{1+p}}{t^p}),
\end{align}
where $r = |x-y|$ is the distance between sites and $\Lambda$ may in some cases be seen as a quantum extension of the classical Lyapunov exponent \cite{Sbao20,Smal16,Shas17,Sroz17,Sgar18,Sxu20}. In this regard, the speed of the wave front $v$ effectively defines a causal light cone, and confines the \enquote{connected} region. In chaotic systems $v$ is usually referred to as the butterfly velocity. The parameter $p$ is model-dependent and controls the broadening of the wave front. 
For a given time $t$, the correlator thus decays exponentially in space. For sites located beyond the light cone $r > vt$, there is also an exponential decay of $C(r,t)$.
Note that the reverse statement is also true: Lieb-Robinson type bounds imply locality of interactions \cite{Swil22}. 

\begin{figure}[t]
    \centering
    \begin{tikzpicture}
        \node (a) [label={[label distance=-0.2 cm]143: \textbf{(a)}}] at (0,0) {\includegraphics{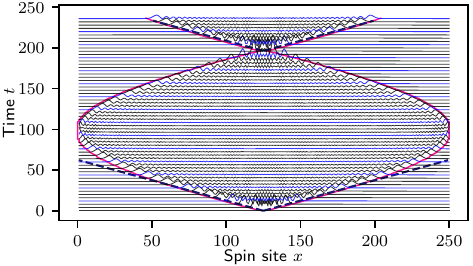}};	
        \node (b) [label={[label distance=-0.2 cm]143: \textbf{(b)}}] at (8.3,0) {\includegraphics{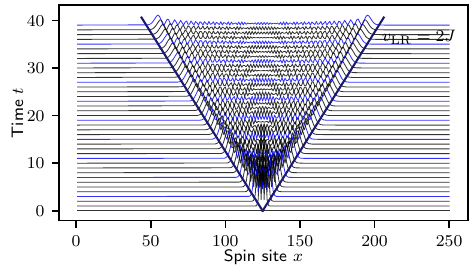}}; 
        \draw (-2,2.5) node {$C(x,N/2,t)$};
        \draw (4.15,3) node {\textsf{Perfect state transfer couplings $J_k = \frac{\lambda}{4}\sqrt{k\cdot(N-k)}$}};
    \end{tikzpicture}
    \caption{Correlator $C(x,N/2,t)$ (in arbitrary units) for a chain of $N=250$ qubits. The average is taken over the infinite temperature maximally mixed state in \cref{eq:C}.
    The function $C(x,N/2,t)$ is shown for discrete times and the lines are offset in $y$-direction \cite{Slin18}. Every fourth line is highlighted in blue for better visibility.
    (a) At early times, the out-of-time-order correlator decays to zero beyond a causally connected region restricted by the finite speed of information propagation, the Lieb-Robinson velocity, $v_\m{LR} = 2J$, which effectively defines a light-cone like structure. The pink line indicates the first occurrence of nonzero correlator $C(x,y,t)$. (b) The boundary deviates from the light cone at later times and operator spreading slows down. Eventually the boundaries are reached at $t = \tau_\m{u}/2 \approx 98$ where the causal region shrinks again to a point at $t = \tau_\m{u}$. Sites once causally connected become disconnected again. This due to high degree of symmetry of the Hamiltonian which imposes symmetries on the correlator, cf. \cref{eq:symtx,eq:symt}.}
    \label{fig:thermal}
\end{figure}

\begin{figure}[t]
    \centering
    \begin{tikzpicture}
        \node (a) [label={[label distance=-0.2 cm]143: \textbf{(a)}}] at (0,0) {\includegraphics{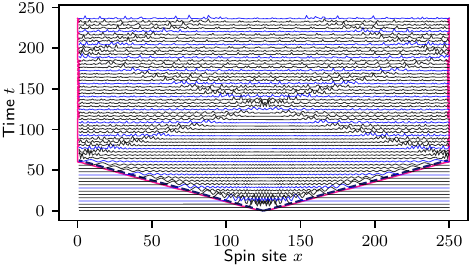}};	
        \node (b) [label={[label distance=-0.2 cm]143: \textbf{(b)}}] at (8.3,0) {\includegraphics{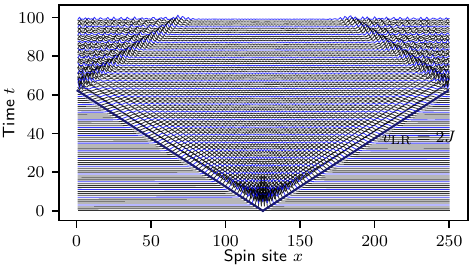}}; 
        \draw (-2,2.5) node {$C(x,N/2,t)$};
        \draw (4.15,3) node {\textsf{Uniform couplings $J_k = J$}};
    \end{tikzpicture}
    \caption{Correlator $C(x,N/2,t)$ (in arbitrary units) for a uniformly coupled chain of $N=250$ qubits with $J_k = J$ $\forall k$ in \cref{seq:H}. The average is taken over the infinite temperature maximally mixed state in \cref{eq:C}. The function
    $C(x,N/2,t)$ is shown for discrete times and the lines are offset in $y$-direction \cite{Slin18}. Every fourth line is highlighted for better visibility.
    (a) Information propagates at the Lieb-Robinson velocity, $v_\m{LR} = 2J$ (dashed line) until the boundaries of the chain are reached. The out-of-time-order correlator is not periodic. Sites that are causally connected remain connected. The pink line indicates the first occurrence of nonzero correlator $C(x,y,t)$, cf. \cref{fig:thermal}a. (b) Short time behavior of (a), the light cone is clearly visible, cf. \cref{fig:thermal}b.}
    \label{fig:thermal-uniform}
\end{figure}

\begin{figure}[t]
    \centering
    \begin{tikzpicture}
        \node (a) [label={[label distance=-0.2 cm]143: \textbf{(a)}}] at (0,0) {\includegraphics{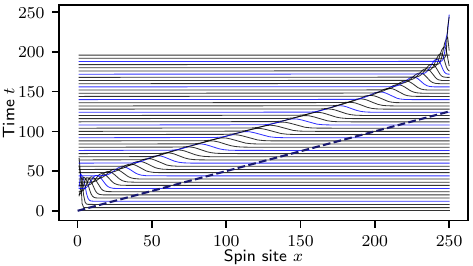}};	
        \node (b) [label={[label distance=-0.2 cm]143: \textbf{(b)}}] at (8.3,0) {\includegraphics{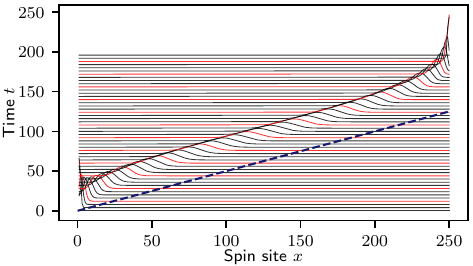}}; 
        \node (c) [label={[label distance=-0.2 cm]143: \textbf{(c)}}] at (0,-4.7) {\includegraphics{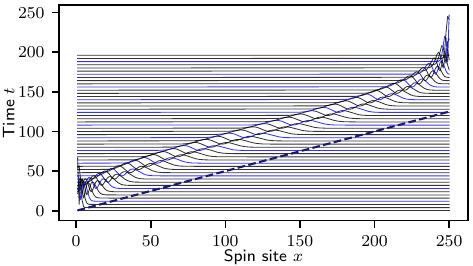}};	
        \node (d) [label={[label distance=-0.2 cm]143: \textbf{(d)}}] at (8.3,-4.7) {\includegraphics{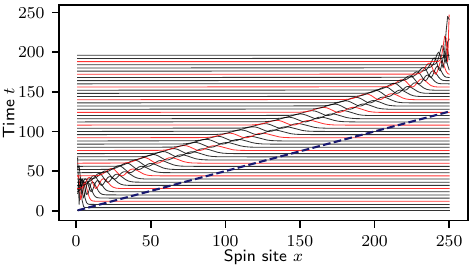}}; 
        \node (e) [label={[label distance=-0.2 cm]143: \textbf{(e)}}] at (0,-9.4) {\includegraphics{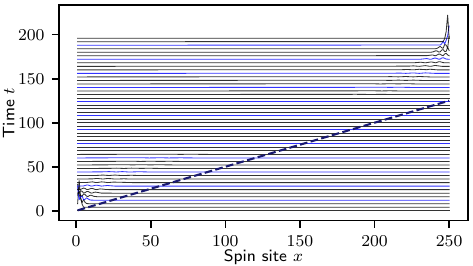}};	
        \node (f) [label={[label distance=-0.2 cm]143: \textbf{(f)}}] at (8.3,-9.4) {\includegraphics{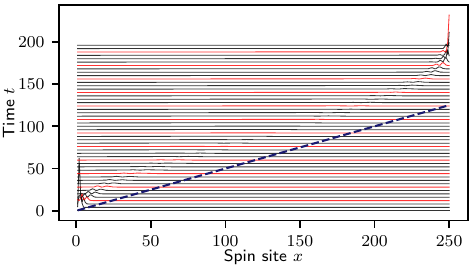}}; 
        \draw (-2,2.5) node {$C(x,1,t)$};
        \draw (-2,-2.2) node {$C(x,2,t)$};
        \draw (-2,-6.9) node {$C(x,3,t)$};
        \draw (0,2.5) node {\textsf{uncorrelated}};
        \draw (8.3,2.5) node {\textsf{correlated}};
        \draw (4.15,3) node {\textsf{Perfect state transfer couplings $J_k = \frac{\lambda}{4}\sqrt{k\cdot(N-k)}$}};
    \end{tikzpicture}
    \caption{Correlator $C(x,y,t)$ (in arbitrary units) for the correlated (left, blue) and uncorrelated (right, red) initial $X$ state for reference spins at different sites $y$. $C(x,y,t)$ is shown for discrete times and the lines are offset in $y$-direction \cite{Slin18}. Every fourth line is highlighted for better visibility. (a),(b) The reference spin is at the first site $y=1$. On each site $x$, the correlator $C(x,y,t)$ is nonzero only for a short time and has one peak. The uncorrelated and correlated correlators coincide. (c),(d) The reference spin is at the second site $y=2$. There are two peaks in $C(x,y,t)$ at each site. The uncorrelated and correlated correlators coincide. (e),(f) The reference spin is at the third site $j=3$. There are three peaks in $C(x,y,t)$ at each site. This situation corresponds to the out-of-time-order correlator depicted in the main text in Fig.~1d. The straight dashed line indicates the Lieb-Robinson velocity for this model given by $v_\m{LR} = 2J$. The upper bound is attained near the center of the chain where the out-of-time-order correlator runs parallel to $v_\m{LR}t$, cf. \cref{sec:LR}.}
    \label{fig:X}
\end{figure}
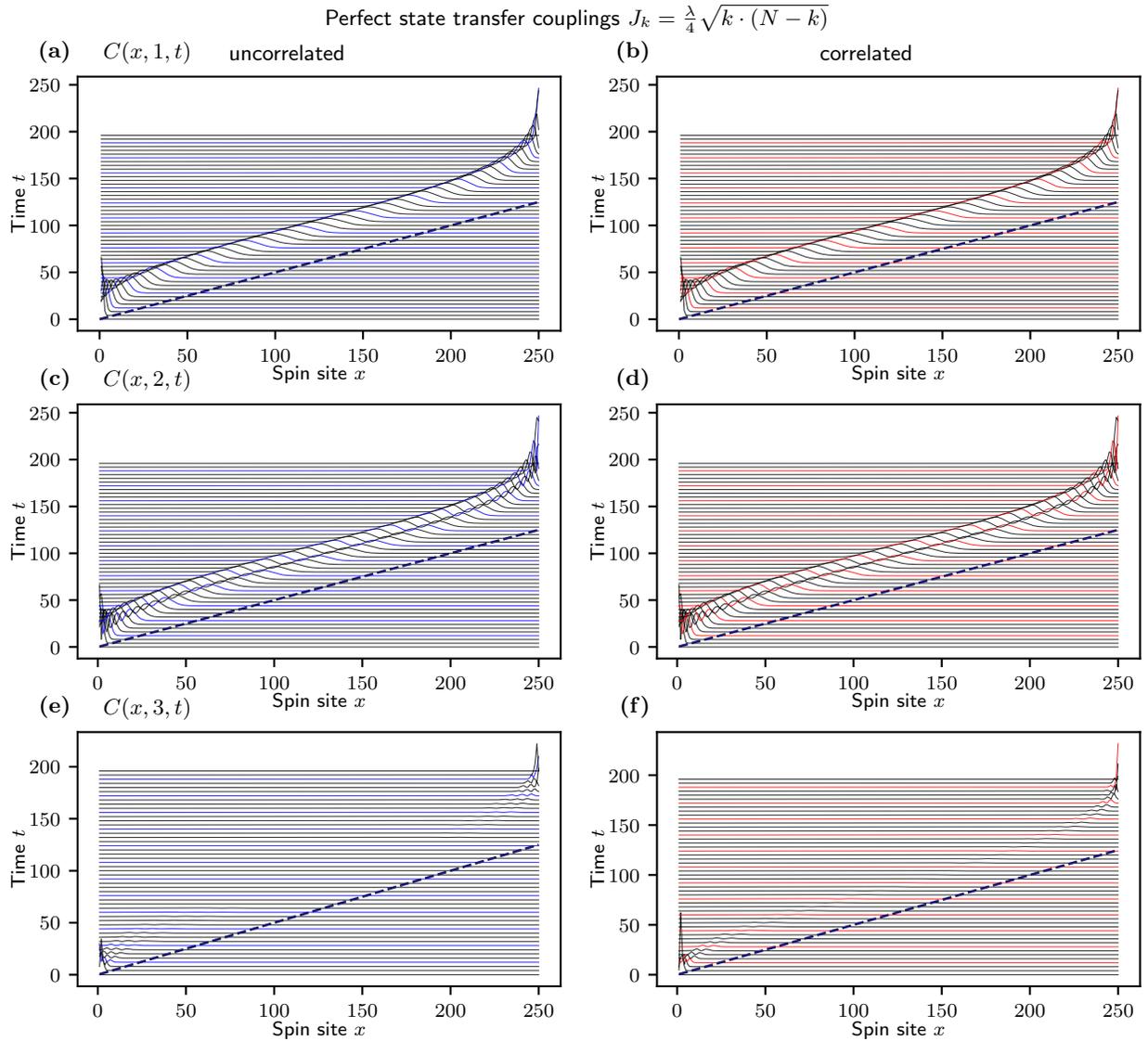

\begin{figure}[t]
    \centering
    \begin{tikzpicture}
        \node (a) [label={[label distance=-0.2 cm]143: \textbf{(a)}}] at (0,0) {\includegraphics{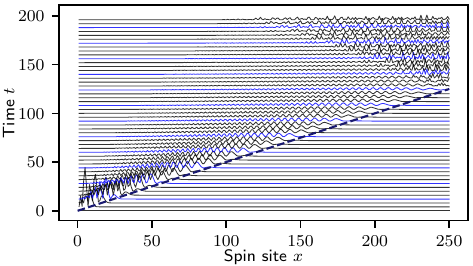}};	
        \node (b) [label={[label distance=-0.2 cm]143: \textbf{(b)}}] at (8.3,0) {\includegraphics{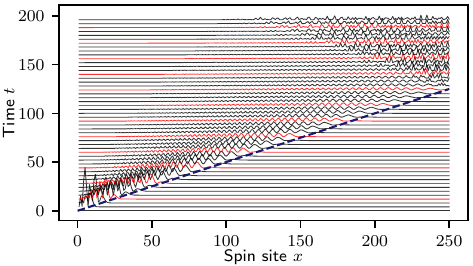}}; 
        \node (c) [label={[label distance=-0.2 cm]143: \textbf{(c)}}] at (0,-4.7) {\includegraphics{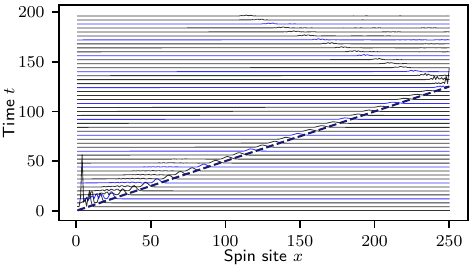}};	
        \node (d) [label={[label distance=-0.2 cm]143: \textbf{(d)}}] at (8.3,-4.7) {\includegraphics{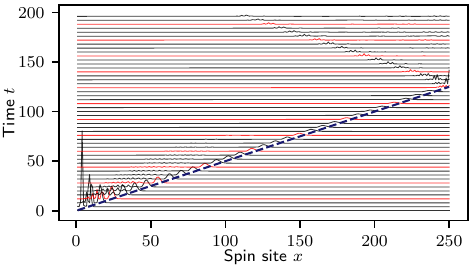}}; 
        \node (e) [label={[label distance=-0.2 cm]143: \textbf{(e)}}] at (0,-9.4) {\includegraphics{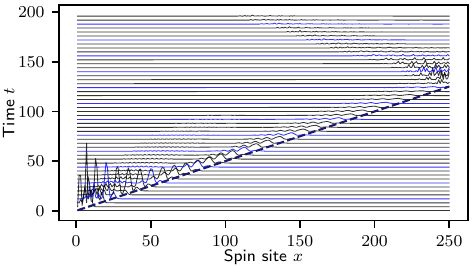}};	
        \node (f) [label={[label distance=-0.2 cm]143: \textbf{(f)}}] at (8.3,-9.4) {\includegraphics{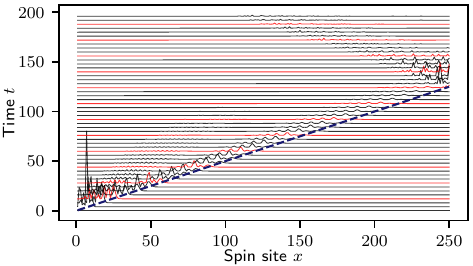}}; 
        \draw (-2,2.5) node {$C(x,1,t)$};
        \draw (-2,-2.2) node {$C(x,2,t)$};
        \draw (-2,-6.9) node {$C(x,3,t)$};
        \draw (0,2.5) node {\textsf{uncorrelated}};
        \draw (8.3,2.5) node {\textsf{correlated}};
        \draw (4.15,3) node {\textsf{Uniform couplings $J_k = J$}};
    \end{tikzpicture}
    \caption{Correlator $C(x,y,t)$ (in arbitrary units) for the correlated (left, blue) and uncorrelated (right, red) initial $X$ state for reference spins at different sites $y$ in the uniformly coupled spin chain with $(\omega_k = 1,J_k = J)$.
    (a),(b) The reference spin is at site $y = 1$; the out-of-time-order correlators coincide.
    (c),(d) The reference spin is at site $y = 2$; the out-of-time-order correlator is visibly enhanced near the left end of the chain for the correlated initial state. The magnitude of the wavefront decays quickly and the effect of the correlations gets less pronounced with the distance $x$. 
    (e),(f) The reference spin is at site $y = 3$. Correlations have a notable impact on the out-of-time-order correlator only up to a moderate distance in $x$.  
    The straight dashed line indicates the Lieb-Robinson velocity for this model given by $v_\m{LR} = 2J$. Contrary to the perfect state transfer model, the bound is attained at all times and information spreads linearly with $v_\m{LR}$. The information transfer is however incomplete across the $N = 250$ spins because information gets scrambled, i.e. the function $C(x,y,t)$ is different from zero at least in parts of the causal region.}
    \label{fig:X-uniform}
\end{figure}

\subsection{Symmetries}
We now proceed by investigating the symmetry properties of the out-of-time-order correlator in time and space.
For an arbitrary uncorrelated initial condition where 
\begin{align}
    Z(0) = \m{diag}(p_1(0),p_2(0),\ldots,p_N(0)),
\end{align}
(cf. \cref{eq:initial-state}), the evolution of the populations will be of the form 
\begin{align}
    p_k(t) = \sum_{m=1}^N |U_{km}|^2.
\end{align}
From the symmetry relations of the Wigner $d$-matrix elements \eqref{23}, it follows that 
\begin{align}
    (-1)^{N-k}U_{km}(t) = U_{N-k+1,l}(T-t)
    \label{eq:uni}
\end{align}
and thus $|U_{km}|^2(t) = |U_{N-k+1,m}|^2(T-t)$.
This  implies that spins opposite to each other undergo time reversed dynamics, i.e. 
\begin{align}
    \langle \sigma^z_k\rangle(t) = \langle \sigma^z_{N+1-k} \rangle(T-t). 
\end{align}
In order to analyze the properties of the out-of-time-order correlator, we need to investigate the Heisenberg operators $\sigma^z_k(t)$.
The Pauli operators in the single excitation sector, or the Jordan-Wigner transformation respectively, are given by
\begin{align}
    \sigma^z_k(0) \sim 2\dyad{k}-\mathds{1},
\end{align}
where $\dyad{k}$ projects onto the $k$th element and $\sim$ indicates the representation in single excitation sector or the Jordan-Wigner transformation respectively.
The corresponding Heisenberg operators evolve according to 
\begin{align}
    \sigma^z_k(t) \sim 2U^\dagger(t) \dyad{k} U(t) - \mathds{1}
    &= 2\sum_{m,n=1}^N U^\ast_{km}(t)U_{kn}(t) \dyad{m}{n}-\mathds{1},\\
    &\overset{\eqref{eq:uni}}{=} 2\sum_{m,n=1}^N U^\ast_{N-k+1,m}(t-T)U_{N-k+1,n}(t-T) \dyad{m}{n}-\mathds{1},\\
    &\sim \sigma^z_{N+1-k}(T-t).
\end{align}
This immediately implies that the out-of-time-order correlator has the same time-reversal mirror symmetry
\begin{align}
    F(x,y,t) = F(N-x+1,N-y+1,t-T).
    \label{eq:symtx}
\end{align}
This is true irrespective of the initial condition.
Again invoking symmetry relations of the Wigner $d$-matrix, we further obtain 
\begin{align}
    U_{kl}(T-t) = U_{k,N-l+1}(t) (-1)^{l-1}
\end{align}
and therefore, proceeding as before, we obtain
\begin{align}
    \sigma^z_k(T-t) &\sim 2\sum_{m,n=1}^N U^\ast_{km}(t)U_{kn}(t) \dyad{m}{n}-\mathds{1},\\
    &= 2\sum_{m,n=1}^N U^\ast_{k,N-m+1}(t-T)U_{k,N-n+1}(t-T) \dyad{m}{n}-\mathds{1},\\
    &\sim \sigma^z_{k}(t).
\end{align}
For symmetric initial states this implies that the out-of-time-order correlator has the additional symmetry 
\begin{align}
    F(x,y,t) = F(x,y,T-t).
    \label{eq:symt}
\end{align}

Depending on the (initial) state $\rho$, over which the quantum mechanical average is performed, some of those symmetries might be absent.
\Cref{fig:thermal} shows the out-of-time-order correlator for the perfect state transfer Hamiltonian \eqref{eq:H} for a chain of length $N = 250$ with the reference site in the middle of the chain, $y = N/2 = 125$.
Here, the quantum mechanical average is taken over the completely mixed state $\rho = \mathds{1}/d$, so both \cref{eq:symtx,eq:symt} are valid, causing the high symmetry of the out-of-time-order correlator.
For small times (compared to the length of the chain), the out-of-time-order correlator spreads linearly (cf. \cref{sec:LR}) with the butterfly velocity of a uniformly coupled chain, $v_\m{LR} = 2J$, \cref{fig:thermal}b.
Considering the full perfect state transfer time up to $\tau_\m{u} \approx 200$ reveals how the engineered couplings conspire to perfectly invert the dynamics.
After the initial linear spread, the out-of-time-order correlator bends and reaches the boundaries at $\tau_\m{u}/2 \approx 98$ where the causal region starts to shrink to a point again at the final time $\tau_\m{u}$.
The system size is fundamentally encoded into the Hamiltonian and therefore the system knows its size beforehand, explaining the curving of the out-of-time-order correlator near the boundaries.

These features are specific to the particular form of the interactions and the resulting symmetries and are thus not expected generically.
To highlight these special properties, we directly compare the behavior to a system with uniform interactions while otherwise keeping the same parameters. 
\Cref{fig:thermal-uniform} displays snapshots of the out-of-time-order correlator at the same times as in \cref{fig:thermal}.
Since now the system is agnostic to its size, the out-of-time-order correlator exhibits a linear growth at the butterfly velocity throughout until hitting the boundaries. 
The operators hence spread to the boundaries faster than with engineered couplings but information is transmitted only incompletely.
Once a region is causally connected it remains connected and after reflecting at the boundaries information gets scrambled all across the chain and becomes practically inaccessible at the receiving end.

In \cref{fig:X} the quantum mechanical average is taken over the local initial $X$ state which is not symmetric.
Therefore only \cref{eq:symtx} applies and the out-of-time-order correlator is invariant only upon inverting space and reversing time.
We compare the uncorrelated (blue) and correlated (red) states and find that the out-of-time-order correlators coincide for $y=1,2$.
The effect of the correlations first becomes apparent for reference spin $y=3$, displayed in \cref{fig:X}e,f (corresponding to to Fig.~1d of the main text). 
Quantum correlations break the time-reversal symmetry, \cref{eq:symt}, and shift the weight towards the wavefront such that more information is transmitted at the same time thereby increasing the energy flow and boosting state transfer.
Because of the small amplitude of the out-of-time-correlator and the large system size, the effect is less pronounced in the illustration in \cref{fig:X}f.
As before, we compare the results to the uniformly coupled chain in \cref{fig:X-uniform} which shows linear growth and scrambling in accordance with \cref{fig:thermal-uniform}.

\subsection{Lieb-Robinson velocity}
\label{sec:LR}
In this section, we show why the perfect state transfer Hamiltonian \cref{seq:H} leads to information propagation in a local region at maximum speed given by the Lieb-Robinson velocity, as observed in \cref{fig:thermal,fig:X}.
The off-diagonal elements of the Hamiltonian in the single excitation sector are given by
\begin{align}
    J_k = \frac{\lambda}{4} \sqrt{k\cdot (N-k)}.
\end{align}
For elements of $J_k$ close to the middle of the chain with $k = N/2 + m$ where $|m| \ll N$, we obtain (with $\lambda = 4J/N$)
\begin{align}
    J_k = 2J\left[1 - \frac{2m^2}{N^2} + \mathcal{O}(m^4)\right].
\end{align}
Thus, in the vicinity of the chain's center, the interaction strengths are constant up to second order in the distance $m$.
Therefore, in a region where the correction is still small, i.e. $2m^2/N^2 \ll 1$, the out-of-time-order correlator exhibits the same linear growth as the uniformly coupled chain.
In \cref{fig:thermal}, the system size is $N=250$ and hence, to a good approximation, the operators spread linearly over the first $m \approx 25$ sites (where $2m^2/N^2 = 0.02$).

\section{No boost for Bell states}
In this section we show that using the canonical Bell states as initial states for the perfect state transfer protocol does not lead to a boost.
The initial two-qubit state $\rho^0_{12} = \rho_X$ \eqref{eq:X}, is separable with zero concurrence for all $\alpha \in [0,1/4]$. 
In particular, entanglement is thus not necessary for faster state transfer.
We observe, in fact, neither a boost in the state transfer nor in the information flow when a maximally entangled Bell state is used. 
The canonical Bell states belong to the class of $X$ states
\begin{align}
    \rho^0_{12} = \rho_X = 
    \begin{pmatrix}
      a & 0 & 0 & w \\
      0 & b & z & 0 \\
      0 & z^\ast & c & 0 \\
      w^\ast & 0 & 0 & d 
    \end{pmatrix},
\end{align} 
where every parameter is a real number equal to either $0$ or $\pm 1/2$.
In this case the correlations take form 
\begin{align}
    \chi^\m{B} = \begin{pmatrix}
        0 & 0 & 0 & w \\
        0 & 0 & z & 0 \\
        0 & z & 0 & 0 \\
        w & 0 & 0 & 0 
      \end{pmatrix},
\end{align}
where either $w = \pm 1/2$ and $z=0$ or vice versa.
Most importantly, Bell states do not satisfy the necessary condition \eqref{eq:condition}, i.e. $[H_\m{int},\chi^\m{B}] = 0$ but leave the out-of-time-order correlator invariant under time-reversal \cref{eq:symt}. 
Therefore Bell states do not affect the dynamics of the system and entanglement is thus neither necessary nor sufficient to boost information transfer.
As demonstrated in Fig.~2 of the main text, maximally entangled states indeed provide the highest enhancement possible but only if the basis of the quantum correlations matches the interaction Hamiltonian in the sense of \cref{seq:condition}.


\begin{thebibliography}{99}
\bibitem{com} The transfer time $\tau_\m{u}$ then scales with the system size $N$, in line with the Lieb-Robinson bound  \cite{gen16}.
\end{thebibliography}
\end{document}